\documentclass[doublecol]{epl2}
\usepackage{amsmath}
\usepackage{epsfig}
\usepackage{graphicx}
\newcommand{\be}{\begin{equation}}
\newcommand{\ee}{\end{equation}}
\newcommand{\bea}{\begin{eqnarray}}
\newcommand{\eea}{\end{eqnarray}}

\renewcommand{\ss}{{\pmb \sigma}}

\newcommand{\rr}{{\bf r}}
\newcommand{\NN}{{\pmb \nabla}}

\newcommand{\vv}{{\bf v}}
\newcommand{\cc}{{\bf c}}
\newcommand{\uu}{{\bf u}}
\newcommand{\FF}{{\bf F}}

\title{Lattice Boltzmann method for inhomogeneous fluids}

\author{S.Melchionna\inst{1} \and U. Marini Bettolo Marconi\inst{2}}

\institute{\inst{1} 
INFM-SOFT, Dipartimento di Fisica,  
Universit\`a di Roma and Istituto
Nazionale di Fisica della Materia,  Piazzale A. Moro 2, 00185,
Roma, Italy\\
\inst{2}INFM-SOFT, Dipartimento di Fisica, 
Universit\`a di Camerino and Istituto
Nazionale di Fisica della Materia, Via Madonna delle Carceri, 62032 ,
Camerino, Italy}


\abstract{
We present a lattice-based numerical method to describe the
non equilibrium behavior of a simple fluid
under non-uniform spatial conditions.
The evolution equation for the one-particle phase-space distribution function
is derived starting from  a  microscopic description of the system.
It involves a series of approximations which are similar to those
employed in theories of inhomogeneous fluids, such as Density 
Functional theory. Among the merits of the present approach:
the possibility to determine the
equation of state of the model, the transport coefficients and to
provide an efficient method of numerical solution
under non-uniform conditions.
The algorithm is tested in a particular non equilibrium situation, namely
the steady flow of a hard-sphere
fluid across a narrow slit. Pronounced non-hydrodynamic oscillations in the
density and velocity profiles are found.}

\begin{document}

\date{\today}
\maketitle

\section{Introduction}
The behavior of a confined fluid can be different from that
of a bulk fluid in many important aspects. First of all the confinement
induces density inhomogeneities which may determine a variety of 
phenomena having no counterpart in bulk systems \cite{lowen}. 
The presence of surfaces, not only alters the average equilibrium
properties of fluids, but also affects their  
time-dependent behavior such as diffusion, momentum and energy transport.
As a result, a great effort is currently devoted to the understanding
of fluid physics at the nanoscale (see \cite{squiresquake} and references therein).

In the last thirty years, a massive effort has been devoted to the
understanding of system at thermodynamic equilibrium and new techniques
have been developed, among these Density Functional theory (DFT) 
being perhaps 
the most versatile \cite{evans}. On the contrary, we do not have a similar control over the
behavior of non equilibrium systems. Dynamical extensions of DFT 
have been introduced and tested successfully in the case of colloidal
suspensions, where the damped character of the dynamics makes the density the only
relevant field \cite{marconitarazona}. 
Instead, in the case of standard fluids one needs to fully account 
for the momentum and energy transport. 
The natural procedure seems therefore
to consider the evolution of both
positions and momenta of the molecules and to derive  
an equation for the phase-space
distribution able to convey the structural information
about the microscopic nature of the fluid. Such a task can be achieved
by using a modified Enskog-Boltzmann approach, which has been 
proposed by different authors \cite{brey,martysshanchen,hedoolen,Davis,guo}.

Our present goal is
to provide a procedure able to describe simultaneously 
the discrete nature of fluids
and the transport properties at interfaces and in confining geometries.
To this purpose, we briefly recall that 
the Lattice Boltzmann (LB) method represents a very efficient scheme
derived from kinetic theory to simulate fluid flows and transport phenomena
\cite{LBgeneral0,LBgeneral}.  
Being based on the continuum Boltzmann equation, it
accounts for a faithful representation of the macroscopic hydrodynamic
behavior.  The original idea of LB is to model fluid flows by
simplified kinetic equations, the Bhatnagar-Gross-Krook (BGK) relaxation 
operator being a popular choice to simulate the Navier-Stokes evolution at macroscopic level
\cite{LBgeneral0,higuerasuccibenzi}.  The LB of McNamara and Zanetti
\cite{McNamaraZanetti} averages the microdynamics by solving the
kinetic equation of the particle distribution instead of tracking the
motion of each particle.  Whereas the original formulation was
designed to describe lattice gas of particles and the
attention was focused on large scale properties, modern
versions of the LB aim to incorporate a more realistic behavior
of the fluid. In particular, since the pioneering work of Shan and
Chen \cite{shanchen,Succi} there has been a large effort to include 
the effect of
microscopic interactions between fluid molecules into the
description \cite{swiftosborneyeomans,martys,martysshanchen,hedoolen,guo}.  
Such interactions have been accounted for by considering
the intermolecular force field at a particular point in the fluid.
This modification is sufficient to describe non-ideal gas behavior
such as phase coexistence and phase separation and various surface
phenomena, but  
it is still unsatisfactory since it does not allow for a systematic
prediction of the macroscopic properties starting from the microscopic
level. 

In other approaches, in order to circumvent such a difficulty 
the collision terms have been dealt with explicitly, but
at the price of introducting gradients of the macroscopic fields 
\cite{martysshanchen,hedoolen,guo}.

Inspired by recent work on inhomogeneous fluids 
\cite{tarazonamarconi,marconimelchionna}, 
we propose an alternative scheme to evaluate the interaction terms involved in the
kinetic equation. As a result, we obtain evolution equations for the 
distribution function whose structure is very similar to that employed 
in Density Functional theory, without the explicit evaluation of
gradients of macroscopic fields.
We believe that the strategy of using coarse-grained 
quantities instead of gradients of the relevant fields may give 
a better representation of fluids confined to narrow systems, as we
have learned from the study of equilibrium fluids, where gradient expansions
usually give poorer results than non-local density functionals.

In addition, the theory allows to compute equilibrium quantities, such as
the surface tension, together with accurate estimates of the bulk transport properties. 



\section{Theory}

In general, the intermolecular potential of a simple fluid can be
decomposed into a repulsive, responsible for the 
microscopic structure at high packing fraction,
and an attractive contribution, which
plays a major role in determining the thermodynamic properties.
However, in the present paper, which serves as to introduce the method,
we confine ourselves to discuss a harshly repulsive fluid, the hard sphere
system. 

The evolution of the 
phase-space one-particle distribution $f(\rr,\vv,t)$ 
is represented by the following transport equation:
\begin{eqnarray}
\partial_{t}f(\rr,\vv,t) +\vv\cdot\NN f(\rr,\vv,t)
+\frac{\FF_e(\rr)}{m}\cdot\NN_v f(\rr,\vv,t) = \nonumber \\
\Omega[f](\rr,\vv,t)
\label{uno}
\end{eqnarray}
where
$\FF_e$ is an external force
and $\Omega[f]$  represents the effect of the
interactions among the fluid particles which we describe via the 
revised Enskog collision operator \cite{vanbeijerenernst}
\bea
&& \Omega[f]
= \sigma^{d-1}\int d\vv_2\int 
d\hat{\ss}\Theta(\hat{\ss}\cdot \vv_{12}) (\hat{\ss} 
\cdot \vv_{12})\times\nonumber\\
&& \{ g_2(\rr,\rr-\ss,t) f (\rr,\vv_1',t)f (\rr-\ss,\vv_2',t)  -
\nonumber \\
&& g_2(\rr,\rr+\ss,t)f(\rr,\vv_1,t)f(\rr+\ss,\vv_2,t)\}
\label{collision}
\eea
where $\vv_1'$ and $\vv_2'$ are scattered velocities determined 
from $\vv_1'=\vv_1-(\hat\ss\cdot\vv_{12})\hat\ss$ 
and $\vv_2'=\vv_2+(\hat\ss\cdot\vv_{12})\hat\ss$, $\sigma$ is the
hard-sphere diameter, $\hat\ss$ is the unit vector directed from one particle to another,
and $g_2(\rr_1,\rr_2|n)$ is the positional pair correlation function.

Equation \eqref{uno}, together with  \eqref{collision},
represents a nonlinear evolution equation for the one-particle 
distribution function, $f(\rr,\vv,t)$, and could
in principle be solved numerically
by brute force,
or by considering a suitable truncation of the open hierarchy of 
equations for the moments of the
distribution function. However, since one is chiefly interested in 
the evolution of
the hydrodynamic moments of the distribution function
$ n(\rr,t)\equiv \int d\vv f(\rr,\vv,t) $,
$ n(\rr,t)\uu(\rr,t)\equiv\int d\vv \vv f(\rr,\vv,t) $
and 
$\frac{3}{2}n(\rr,t)k_B T(\rr,t)\equiv\frac{m}{2}\int d\vv (\vv-\uu)^2 f(\rr,\vv,t)$,
it is possible to further simplify the form of $\Omega$ without altering
the local transfer of momentum and energy.
To this purpose, following Dufty, Santos and Brey (DSB) \cite{brey}, 
we replace \eqref{uno} and \eqref{collision} by
a simpler equation, where
the complicated interaction between non hydrodynamic modes
is approximated via a BGK-like relaxation term
$-\nu_0 [ f(\rr,\vv,t)-n(\rr,t)\phi_M(\vv)]$, 
$\nu_0$ being a phenomenological collision frequency, chosen as to reproduce the 
kinetic contribution to the viscosity. The DSB equation can be cast in the form:
\bea
\partial_{t}f(\rr,\vv,t) &+&\vv\cdot\NN f(\rr,\vv,t)
+\frac{\FF_e(\rr)}{m}\cdot\NN_v f(\rr,\vv,t)=
\nonumber\\
&-& 
\nu_0 [ f(\rr,\vv,t)-n(\rr,t)\phi_M(\rr,\vv,t)]\nonumber\\
&+& m \beta \phi_M(\rr,\vv,t)\left\{(\vv-\uu)\cdot {\bf C}^{(1)}(\rr,t)
\right.
\nonumber\\
&+& 
\left.
(\frac{m\beta (\vv-\uu)^2}{d} -1 ) C^{(2)}(\rr,t)
\right\}
\label{brey}
\eea
where  $\beta=1/k_B T(\rr,t)$ and
$ \phi_M(\rr,\vv,t)=[\frac{m}{2\pi k_B T(\rr,t)}]^{3/2}\exp
\Bigl(-\frac{m(\vv-\uu)^2}{2 k_B T(\rr,t)} \Bigl)$.
Moreover,
\bea
C^{(1)}_\mu(\rr,t) &=& \int d\vv v_\mu \Omega 
\equiv - \frac{1}{m} \sum_\nu \nabla_\nu P^c_{\nu\mu}(\rr,t)
\label{gradient}
\eea
where $P_{\nu\mu}^c$ is the collisional transfer part of the pressure tensor, and
\bea
C^{(2)}(\rr,t) &=& \frac{1}{2}\int d\vv \sum_\nu (v_\nu - u_\nu)^2 \Omega
\nonumber \\
&\equiv& - \frac{1}{m} \sum_\nu [\nabla_\nu q_\nu ^c(\rr,t)
+\sum_\mu P_{\nu\mu}^c(\rr,t) \nabla_\nu u_\mu(\rr,t)]
\nonumber \\
\label{gradienth}
\eea
arises from the collisional contribution to the heat flux, ${\bf q}^c$.

Equation \eqref{brey} reproduces the correct form of the hydrodynamic
equations, but treats in an approximate fashion, viz. within a relaxation time
approximation, the higher velocity moments contributing to
$f(\rr,\vv,t)$  and associated with kinetic
modes \cite{brey}.

The specific forms of ${\bf C}^{(1)}$ and $C^{(2)}$ 
needed to render feasible the method will be given below.
Before doing that, we notice
that formula \eqref{gradient} establishes a connection 
between the present method and Density Functional theory. In fact, 
in the case of vanishing velocity and temperature gradients,
eq. \eqref{gradient} can be expressed with the help
of the  excess Helmholtz free energy,
${\cal F}_{exc}[n]$, as ${\bf C}^{(1)}(\rr,t) = 
- \frac{1}{m} n(\rr,t)\NN \frac{\delta {\cal F}_{exc}[n]}{\delta n(\rr,t)}$.  
On the other hand, the term $C^{(2)}(\rr,t)$ describes non isothermal processes 
and has no counterpart within the standard DFT formalism.
Under equilibrium conditions, i.e. in the absence
of velocity gradients and thermal gradients one can derive the following 
equation of state  with the bulk pressure given by
\be p_{bulk}=\frac{1}{d}\sum_{\nu=1}^d
\Bigl[P_{\nu\nu}^{kin}+ P^{c}_{\nu\nu} \Bigr]_{bulk}=
k_B T\Bigl[n_b+ \frac{2\pi}{3}n_b^2\sigma^3 g_2(\sigma)\Bigr]
\ee
where we set
the kinetic contribution to the pressure tensor 
$P_{\nu\mu}^{kin}=m\int d \vv (v-u)_\nu(v-u)_\mu f(\rr,\vv,t)$ to be
diagonal and equal to the ideal gas value.
In the general non equilibrium case, we consider 
eq. \eqref{brey} by taking into account
the full contribution to the pressure tensor stemming, e.g., 
from shearing perturbations.


\section{Methods and Results}

We shall solve eq. \eqref{brey} by means of the
lattice Boltzmann method \cite{LBgeneral} which has the advantage of
being very efficient and robust and
accomodates in a natural fashion informations about the microscopic model.
We restrict our attention to the isothermal situation.

The original LB method can be viewed as a discretization procedure in velocity 
space of a kinetic equation \cite{shanhe,martysshanchen,heluo}. We will employ the popular D3Q19 lattice, 
constituted by a set of $19$ populations. The continuous velocity $\vv$ is
replaced by a set of discrete velocities $\cc_i$, with $i=0,18$, which
are vectors which connect neighboring mesh points $\rr$ on a lattice
and where the null vector $\cc_0$ accounts for particles at rest.
Accordingly, the distribution function is replaced by
an array of $19$ populations, $f(\rr,\vv,t)\rightarrow f_i(\rr,t)$. 

The systematic procedure to discretize the kinetic equation is based on
expanding the distribution in a finite set of Hermite polynomial
$f(\rr,\vv,t)=\omega(v)\sum_{l=0}^K
\Phi^{(l)}_{\underline\alpha}(\rr,t)h^{(l)}_{\underline\alpha}(\vv) 
/v_T^{2l} 2l!
$ 
where $\omega(v) = (2\pi v_T^2)^{-2d}e^{-v^2/2v_T^2} $,
$v_T=\sqrt{k_BT/m}$
and by considering a Gauss-Hermite quadrature of order $2G$ to evaluate the
integrals of the type
\bea
\int d\vv v^{\alpha_1}...v^{\alpha_M} &f(\rr,\vv,t)& \nonumber \\
&=& \sum_{i=0}^G w_i c_i^{\alpha_1}...c_i^{\alpha_M} f(\rr,\cc_i,t)/\omega(c_i)
\nonumber \\
&=& \sum_{i=0}^G c_i^{\alpha_1}...c_i^{\alpha_M} f_i(\rr,t)
\eea
where $w_i$ are quadrature weights and $f_i(\rr,t) \equiv w_i f(\rr,\cc_i,t)/\omega(c_i)$.
Therefore, the kinetic moments are evaluated via
\bea
\Phi^{(l)}_{\underline\alpha}(\rr,t) = 
\sum_{i=0}^{K}f_{i}(\rr,t)h^{(l)}_{\underline\alpha}(\cc_i)
\eea
Analogously, we expand the collision operator on the finite Hermite set,
$$\Omega(\rr,\vv,t)=\omega(v)\sum_{l=0}^K
\frac{1}{v_T^{2l}2l!}
C^{(l)}_{\underline\alpha}(\rr,t)h^{(l)}_{\underline\alpha}(\vv)
$$
and evaluate its moments via
\bea
C^{(l)}_{\underline\alpha}(\rr,t) = \int d\vv 
\Omega(\rr,\vv,t) h^{(l)}_{\underline\alpha}(\vv)
\eea

The propagation of the populations is achieved via a time discretization
to first order and a forward Euler update, 
\bea
f_i(\rr+\cc_i,t+1) - f_i(\rr,t) = w_i\sum_{l=0}^K \frac{1}{v_T^{2l}l!} 
C_{\underline \alpha}^{(l)}(\rr,t) h_{\underline \alpha}^{(l)} (\cc_i)
\eea

Concerning the BGK term appearing in eq. \eqref{brey}, standard calculations lead to the
following expression for the local equilibrium
\bea
n(\rr,t)\Phi_M(\cc_i) && \nonumber \\
&=& 
w_i n(\rr,t)\left[
1 + \frac{1}{v_T^2} \sum_\nu c_{i\nu} u_\nu(\rr,t) 
\right.
\nonumber \\
&+&
\left.
\frac{1}{2v_T^4}\sum_{\nu,\mu} (c_{i\mu} c_{i\nu} - v_T^2 \delta_{\mu\nu}) u_\nu(\rr,t) u_\mu(\rr,t)
\right]
\nonumber \\
\eea
which is tantamount to a low-Mach ($O[Ma^3]$) expansion.

In order to obtain a practical scheme we have to specify the explicit
form of the collisional contributions to the pressure and to the heat flux.
This is done by considering the fact that in a hard-sphere fluid, momentum and energy
fluxes can be transferred instantaneously even when
the velocity distribution function
has a Maxwellian form, provided it peaks at
the local hydrodynamic velocity with a
spread determined by the local temperature.
The explicit expression for the collisional contributions then yields
$C^{(0)}_i = 0$, fulfilling mass conservation, and
\bea
C^{(1)}_\nu(\rr,t) &=& - v_T^2
\int d^2\sigma \hat{\sigma}_\nu g_2(\rr,\rr+\ss | n)
n(\rr,t) n(\rr+\ss,t) \nonumber \\
&\times& \left\{
 1 - \frac{2}{v_T \sqrt{\pi}}
\sum_\nu \hat\sigma_\nu 
\left[ u_\nu(\rr+\ss,t) - u_\nu(\rr,t) \right] 
\right. 
\nonumber \\
&&+ 
\left. 
\frac{k_B[T(\rr+\hat\ss,t)-T(\rr,t)]}{2mv_T^2}
\right\}
\label{LBHS}
\eea
The next $C^{(2)}$ term, governing energy transport, can be derived 
explicitly and reads
\bea
C^{(2)}(\rr,t) &=& v_T^2  \sigma^2
\int d^2 \hat{\sigma}
g_{2}(\rr,\rr+\ss|n)n(\rr,t)n(\rr+\ss,t)
\nonumber\\
&\times&
\Bigr\{
-\sum_{\nu=1}^3\hat{\sigma}_{\nu}
\frac{[u_\nu(\rr+\ss,t) - u_\nu(\rr,t)]}{2}
\nonumber\\
&&
+\frac{1}{\sqrt \pi}
\frac{k_B[T(\rr+\ss,t)-T(\rr,t)]}{mv_T}
\Bigr\}
\label{C2eq}
\eea

Having, now, an explicit representation for $C^{(1)}_\nu$ and $C^{(2)}$
in terms of the local values of the five hydrodynamic fields we can
solve by iteration the equation of evolution for $f(\rr,\vv,t)$, compute
the new values of the fields and proceed.
Notice that in order to extract the momentum flux and the heat flux
one has to use formulae  \eqref{gradient} and \eqref{gradienth}

The inclusion of energy transport in LB is, however,
conditioned by the stability of the numerical method. 
In fact, it is well known that non-isothermal LB schemes suffer from rather
severe instabilities already at the level of ideal fluids \cite{LBgeneral}. 
In the following, we have thus considered the regime in which thermal gradients
are small, i.e. by taking $T=const$ in eqs. (\ref{LBHS}) and (\ref{C2eq}).

The radial distribution function appearing in \eqref{LBHS} is
constructed according to the following prescription, dating back to
Fischer and Methfessel \cite{fishmet}.  At first, one defines a
coarse-grained density $\bar n (\rr,t)$ via a uniform smearing over a
sphere of radius $\sigma/2$, and the coarse-grained packing fraction
is $\bar \eta (\rr,t) = \pi\sigma^3 \bar n(\rr,t)/6$. Next, the
equilibrium radial distribution function, $g_2 (\rr,\rr+\ss|n)$, is
replaced by the following approximation \cite{hansenmcdonald} $
g_2(\rr,\rr+\ss) \simeq
[(1-\bar\eta(\rr+\ss/2)/2+\bar\eta^2(\rr+\ss/2)/4]/[1-\bar\eta(\rr+\ss/2)]^3$.

In order to evaluate the surface integrals we choose $\sigma$ to be an
even multiple of the lattice spacing and employ a $18$-point
quadrature over a spherical surface \cite{abramovitzstegun}.  With
this choice, the elements arising from $g_2$ and the hydrodynamic
moments are taken from $6$ on-lattice quadrature points while the
elements arising from the remaining $12$ off-lattice points are
constructed via a linear interpolation from the surrounding on-lattice
elements.

We have found that this approximation for the pair correlation function
gives excellent results when the coupling between density and current
in eqs. (\ref{LBHS}) and (\ref{C2eq}) is absent, corresponding to a dynamical DFT treatment.
However, when the coupling is active, correlations are spuriously
enhanced even for the static case.  This problem can be traced back
to the well-known compressibility error intrinsic to the LB
discretization \cite{LBgeneral}, i.e. the fluid current is not rigorously divergence
free in absence of external forcing. In order to alleviate such a
problem, we have introduced in the density-current convolution of eq.(\ref{LBHS})
a space-dependent regularizing factor $(1 - \exp(-|r-r_w|/\sigma))$, 
where $r_w$ is the position of the wall.
An alternative to this intervention could be the use of an Hermite basis set with 
higher components than the one associated to the D3Q19 lattice scheme.

\subsection{Numerical validation}

\begin{figure}[htb]
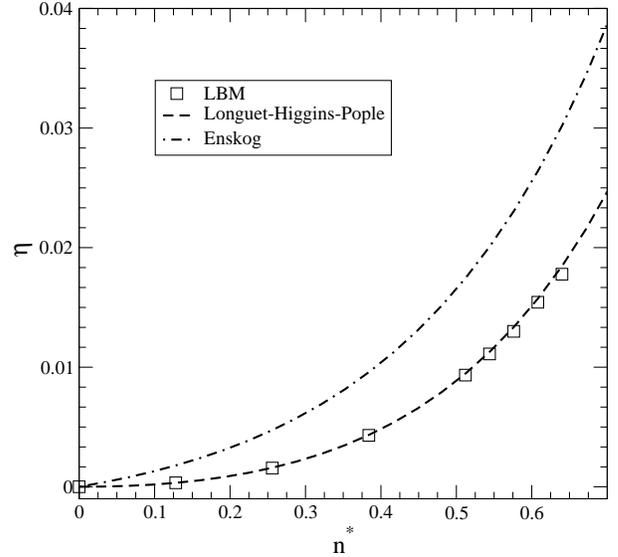

\onefigure[clip,width=0.90\columnwidth]{etavsrho.eps}
\caption{Collisional contribution to the excess part of the bulk shear viscosity 
obtained through: the 
Longuet-Higgins and Pople theory 
($\eta=\frac{4}{15}\sqrt{k_BTm} g_2(\sigma) (n^{\star}_{bulk})^2 \sigma^4$) \cite{Longuet} 
(dashed line),
Enskog theory
($\eta=\frac{5}{16\sigma^2}\sqrt{k_BTm \over \pi} 
(0.8 \frac{2\pi}{3}n^{\star}_{bulk} 
+ 0.7737 g_2(\sigma) \frac{4\pi^2}{9} (n^{\star}_{bulk})^2 )$ \cite{SiguHeyes} 
(dot-dashed line), 
and numerical data obtained from the time decay of a transverse velocity perturbation (squares).
 }
\label{ETA}
\end{figure}

As a first test of our scheme we have determined the shear 
viscosity of a uniform system by performing a linear analysis of
the equations of motions for the hydrodynamic fields (to be published elsewhere), 
obtained from eq.~(\ref{brey}) and eq.~(\ref{LBHS}). It can be shown that the present theory
gives a shear viscosity identical to that predicted by a method
proposed long ago by Longuet-Higgins and Pople (LHP) \cite{Longuet}. In 
fig.\ref{ETA} we display the numerical results 
together with the predictions of the 
LHP theory and the Enskog theory which, as it is well known,
gives a value of $\eta$ larger than that obtained from the LHP route \cite{SiguHeyes}.
 
\begin{figure} [htb]
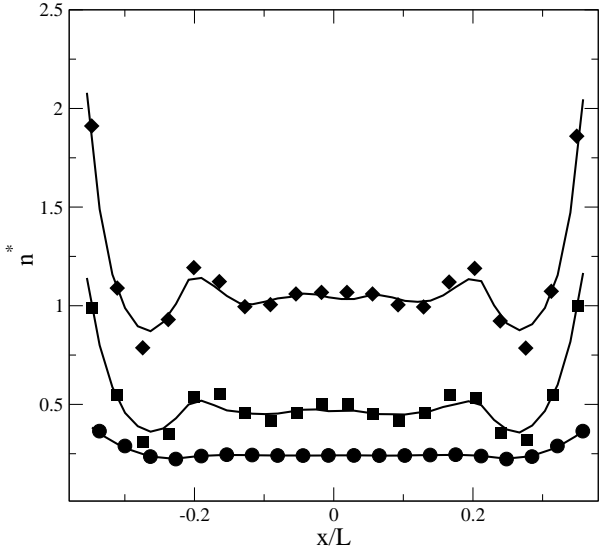

\onefigure[clip,width=0.90\columnwidth]{static.eps}
\caption{Static density profiles in a slit of width $L/\sigma=5$ at 
$n^{\star}_{bulk}=0.256$ (lower curve),
$n^{\star}_{bulk}=0.512$ (mid curve)
and  $n^{\star}_{bulk}=0.609$ (upper curve) as compared to Monte Carlo
simulations. The data at higher density have been shifted by an arbitrary value.}
\label{STATICS}
\end{figure}

Next, we have considered the equilibrium structure of the hard-sphere
fluid.
No-slip boundary conditions are enforced on the
fluid velocity at the wall via a bounce-back method \cite{LBgeneral}.
The corresponding density profiles
are reported in fig.\ref{STATICS} at various values of the
average density together with a comparison with the results of
equilibrium Monte Carlo simulations. 
Clearly, the fluid is more inhomogeneous at higher densities
and displays oscillatory behavior in the regions adjacent to the walls.
These oscillations become less evident toward the center of the slit.
As compared to the exact Monte Carlo solution, the LB data provide slightly
less correlated profiles. It is worth mentioning that the LB solution
is achieved with a CPU effort about $30$ times smaller than needed to
generate converged Monte Carlo data.

\begin{figure} [htb]
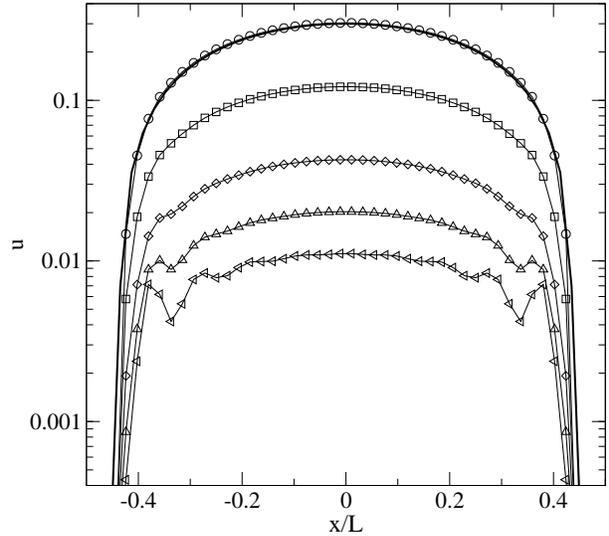

\onefigure[clip,width=0.90\columnwidth]{J.eps}\par
\caption{Poiseuille velocities 
under the influence of the external field $F_e/\sigma^3 = 2.4 10^{-6}$
for a slit of width $L/\sigma = 10$ 
at various bulk densities:
$n^\star_{bulk}=0.064$  (circles), $0.128$  (squares), $0.256$  (diamonds), 
$0.384$  (triangles up), and $0.512$  (triangles left).  }
\label{POISE}
\end{figure}

By considering the fluid flow
induced by the presence of a uniform field, ${\bf F}_e$, parallel
to the walls of the slit, the streaming velocity profiles corresponding
to different values of the bulk density
are shown in fig.\ref{POISE}.
We notice that
as the bulk density increases also the current
profiles display a non-monotonic structure close to the
walls. In addition, 
the average streaming velocity decreases as the density increases as
a consequence of mutual steric hindrance among particles.

\begin{figure} [htb]
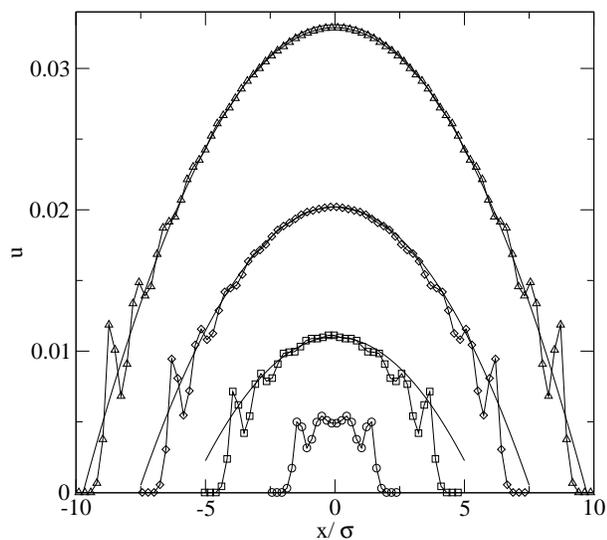

\onefigure[clip,width=0.90\columnwidth]{UvsL.eps}\par
\caption{Poiseuille velocities under the same forcing of fig. \ref{POISE}
for $n^\star_{bulk}=0.512$
at various slit widths:
$L=20\sigma$ (triangles),
$15\sigma$ (diamonds),
$10\sigma$ (squares) and
$5\sigma$ (circles).}
\label{POISE2}
\end{figure}

Finally, in fig.~\ref{POISE2} we have considered the dependence of
the streaming profiles on the width of the slit and, for the sake of comparison,
displayed the results against the  parabolic velocity profiles
{\em \`a la} Poiseuille, 
predicted by the  Navier-Stokes theory. Interestingly,  
the central region does not exhibit significant deviations from 
the quadratic behavior of the classical theory, while a non monotonic
profile next to the walls becomes more pronounced for
the narrower systems \cite{Travis}. 

\section{Conclusions}

In summary, we have presented a theoretical analysis and computational scheme
which bridge hydrodynamics with microscopic structural theories of fluids.
We stress that the present method has been derived starting from a
microscopic model of the fluid and, unlike the very popular and
succesfull Shan-Chen method, truly represents a bottom-up approach
to the description of fluid transport.  

In order to simulate microscopic flows, however, the LB method should be employed with care.
In nanofluidic conditions, the ratio between the mean free path and the characteristic length
(the Knudsen number) can be rather high. On the contrary, the LB method in the current
implementation (with the D3Q19 finite Hermite basis) is designed to deal with low-Knudsen
conditions, namely in the range of validity of the Navier-Stokes equations. Recently, however,
a generalization of the numerical method to deal with high-Knudsen conditions has been  proposed
\cite{shanyuanchen},
that basically extends the Hermite basis up to the third order and removes off-basis contributions.
We believe that such an extension, together with the kinetic approach described in this paper,
would yield a relevant step forward in the study of out-of-equilibrium microscopic flows.

A straighforward
generalization of the present work can accomodate attractive forces,
which are essential ingredients in order to describe multi-phase
behavior.
A detailed description of the calculations reported in the present communication
will be published elsewhere.

\section{Acknowledgments}

U.M.B.M. acknowledges the support of the 
Project Complex Systems and Many-Body Problems
Cofin-MIUR 2005 prot. 2005027808. 

\end{document}